\documentstyle[sprocl]{article}
\begin{document}

\title{RECOVERING RELATIVISTIC NUCLEAR PHENOMENOLOGY 
FROM THE QUARK-MESON COUPLING MODEL}

\author{XUEMIN JIN and B.K. JENNINGS}

\address{TRIUMF 4004 Wesbrook Mall, Vancouver, B.C., Canada V6T 2A3}

\maketitle

\abstracts{The quark-meson coupling (QMC) model for nuclear matter, 
which describes nuclear matter as non-overlapping MIT bags bound by 
the self-consistent exchange of scalar and vector mesons is modified 
by the introduction of a density dependent bag constant. 
It is found that when the bag constant is significantly reduced in nuclear 
medium with respect to its free space value, large canceling isoscalar 
Lorentz scalar and vector potentials for the nucleon in nuclear matter 
emerge naturally. Such potentials are comparable to those suggested by 
relativistic nuclear phenomenology. This suggests that the modification 
of the bag constant in the nuclear medium may play an important role in 
low- and medium-energy nuclear physics.}


While the description of nuclear phenomena has been efficiently
formulated using the hadronic degrees of freedom, new challenges arise
from the observed small but interesting corrections to the standard
hadronic picture such as the EMC effect which reveals the medium
modification of the internal structure of nucleon. To
address these new challenges, it is necessary to build theories that
incorporate quark-gluon degrees of freedom, yet respect the
established theories based on hadronic degrees of freedom.

A few years ago, Guichon \cite{guichon88} proposed a quark-meson
coupling (QMC) model to investigate the direct ``quark effects'' in
nuclei. This model describes nuclear matter as non-overlapping MIT
bags interacting through the self-consistent exchange of mesons in the
mean-field approximation \cite{guichon88,fleck90}. The QMC model, however, 
predicts much smaller scalar and vector potentials for the nucleon than obtained in
relativistic nuclear phenomenology. The latter is a general approach based on
nucleons and mesons which has gained tremendous credibility during last
twenty years \cite{wallace87}. It is known that the large and canceling 
scalar and vector potentials are central to the success of the relativistic 
nuclear phenomenology. 

We observe that the bag constant is held to be at its free space value in the
QMC model. This assumption can be questioned. 
The bag constant in the MIT bag model contributes $\sim 200-300$ MeV
to the nucleon energy and provides the necessary pressure to confine
the quarks.  Thus, the bag constant is an inseparable ingredient of
the bag picture of a nucleon. When a nucleon bag is put into the
nuclear medium, the bag as a whole reacts to the environment. As a
result, the bag constant may be modified.  There is little doubt that
at sufficiently high densities, the bag constant is eventually melted
away and quarks and gluons become the appropriate degrees of freedom.
Therefore, It is reasonable to believe that the bag constant is modified and
decreases as density increases. This physics is obviously bypassed in
the QMC model by the assumption of $B=B_0$ \cite{guichon88,fleck90}.

We modify the QMC model by introducing a density dependent bag constant. 
We model the density dependence of the bag constant in
two ways: one invokes a direct coupling of the bag constant to the
scalar meson field, and the other relates the bag constant to the in-medium nucleon 
mass \cite{jin96}. Both models feature a decreasing bag constant with increasing density.
 We find that the reduction of the bag constant 
in nuclear matter partially offsets the effect of the internal quark structure 
of the nucleon and effectively introduces a new source of attraction. This 
attraction needs to be compensated with additional vector field strength.
The decrease of bag constant also implies the increase of bag radius in nuclear 
matter. This is consistent with the ``swollen'' nucleon picture discussed in the
literature.

When the bag constant is reduced significantly in nuclear matter with
respect to its free-space value, we find that our modified quark-meson 
coupling model predicts large and canceling scalar and vector potentials 
for the nucleon in nuclear matter, which is qualitatively different from 
the prediction of the simple QMC model. These potentials are consistent with those
suggested by the relativistic nuclear phenomenology, implying that the essential 
physics of the relativistic nuclear phenomenology can be recovered. This suggests 
that the drop of the bag constant in nuclear medium may play an important role 
in low- and medium-energy nuclear physics.

\section*{Acknowledgments}
This work was supported by the Natural Sciences and Engineering 
Research Council of Canada.

\end{document}